\documentclass[structabstract]{aa} 
\usepackage{graphicx}
\usepackage{natbib}
\usepackage{url}
\usepackage{amsmath}
\usepackage{subeqnarray}
\usepackage{layouts}
\usepackage{color}
    \definecolor{Blue}{rgb}{0.0,0.0,1.0}
    \definecolor{Red}{rgb}{1.0,0.0,0.0}
    \definecolor{Green}{rgb}{0.0,1.0,0.0}
\newcommand{\be}{\begin{equation}}                                 
\newcommand{\ee}{\end{equation}}                                   
\newcommand{\bea}{\begin{eqnarray}}                                
\newcommand{\eea}{\end{eqnarray}}  
\newcommand{\bi}{\begin{itemize}}                                 
\newcommand{\ei}{\end{itemize}}                                      
\newcommand{\nn}{\nonumber}                                                   
\definecolor{gray}{rgb}{.6,.6,.6}                                  %
\definecolor{green}{rgb}{0,.6,0}                                   %
\definecolor{red}{rgb}{0.6,0,0}                                    %
\newcommand{\xb}{\bar{x}}
\newcommand{\yb}{\bar{y}}
\begin{document}
%-----------------------------------------------------------------------
\title{Spectral signature of oscillating slender tori surrounding\\Kerr black holes}
%
%   \subtitle{}
%-----------------------------------------------------------------------
\author{      F. H. Vincent\inst{1}
%-----------------------------------------------------------------------
\and           G. P. Mazur\inst{1, 2}
%-----------------------------------------------------------------------
\and           O. Straub\inst{3}
%-----------------------------------------------------------------------
\and           M. A. Abramowicz\inst{1, 4, 5}
%-----------------------------------------------------------------------
%\and           M. Johansson\inst{4}
%-----------------------------------------------------------------------
%\and           E. {\v S}ramkov{\'a}\inst{5}
\and           W. Klu\'zniak\inst{1}
%-----------------------------------------------------------------------
\and           G. T{\"o}r{\"o}k\inst{5}
%-----------------------------------------------------------------------
\and           P. Bakala\inst{5}
%-----------------------------------------------------------------------
%-----------------------------------------------------------------------
}
%-----------------------------------------------------------------------
\institute{
%-----------------------------------------------------------------------
              Nicolaus Copernicus Astronomical Center, ul. Bartycka 18, PL-00-716 Warszawa, Poland
	   \\ \email{fvincent@camk.edu.pl}
%-----------------------------------------------------------------------
\and         Faculty of Physics, Warsaw University, ul. Hoza 69, PL-00-681 Warszawa, Poland
                 \\ \email{gmazur@camk.edu.pl}
%-----------------------------------------------------------------------
\and         LUTH, Observatoire de Paris, CNRS, Universit\'e Paris Diderot, 5 place Jules
Janssen, 92190 Meudon, France
                  \\ \email{odele.straub@obspm.fr}
%-----------------------------------------------------------------------
\and          Physics Department, Gothenburg University,
               SE-412-96 G{\"o}teborg, Sweden
                 \\ \email{marek.abramowicz@physics.gu.se}
%-----------------------------------------------------------------------
%\and         Chalmers University of Technology
%                 \\ \email{matzjb@yahoo.se}
%-----------------------------------------------------------------------
\and         Institute of Physics, Faculty of Philosophy and Science, Silesian University in Opava, Bezrucovo nam. 13, CZ-746 01 Opava, Czech Republic
                  \\ \email{gabriel.torok@gmail.com}
                  \\ \email{pavel.bakala@fpf.slu.cz}
                %  \\ \email{sram\_eva@centrum.cz}
%-----------------------------------------------------------------------
}
%-----------------------------------------------------------------------
   \date{Received ; accepted }
%-----------------------------------------------------------------------
%-----------------------------------------------------------------------
\abstract 
  % context heading (optional)
  % {} leave it empty if necessary  
   {Some microquasars exhibit millisecond quasi-periodic oscillations (QPO) that are likely related to phenomena
   occuring in the immediate vicinity of the central black hole.
   Oscillations of accretion tori have been proposed to model these QPOs.}
  % aims heading (mandatory)
   {Here, we aim at determining the observable spectral signature of slender accretion tori
   surrounding Kerr black holes. We analyze the impact of the inclination and spin parameters
   on the power spectra.}
  % methods heading (mandatory)
   {Ray-traced power spectra of slender tori oscillation modes are computed in the Kerr metric.}
  % results heading (mandatory)
   {We show that the power spectral densities of oscillating tori are very sensitive
   to the inclination and spin parameters. This strong dependency of the temporal spectra
   on inclination and spin may lead to observable constraints of these parameters.}%
     % conclusions heading (optional), leave it empty if necessary 
   {This work goes a step further in the analysis of the oscillating torus
   QPO model. It is part of a long-term study that will ultimately lead to
   comparison with observed data.}

%-----------------------------------------------------------------------
%-----------------------------------------------------------------------
   \keywords{Accretion, accretion disks -- Black hole physics -- Relativistic processes
               }
%-----------------------------------------------------------------------
\authorrunning{Vincent et al.}\titlerunning{Oscillating tori for modeling QPOs}
%-----------------------------------------------------------------------
\maketitle
%-----------------------------------------------------------------------

%-----------------------------------------------------------------------

%%%%%%%%%% 	INTRODUCTION	  %%%%%%%%%%%%%

\section{Introduction}
%-----------------------------------------------------------------------
%-----------------------------------------------------------------------

Some microquasars exhibit millisecond high-frequency quasi-periodic oscillations 
(QPO)
characterized by a narrow peak in the source power 
spectrum~\citep[see][for a review]{vanderklis04,remillard06}. 
Their characteristic time scale is of the order of the Keplerian orbital period
at the innermost stable circular orbit of the central $\approx 10\,M_{\odot}$ black hole. It is thus probable
that they are related to strong-field general relativistic effects.

A variety of models have been developed over the years to account for this high-frequency variability.
The nature of high-frequency QPOs and their association with the steep power-law state of luminous accretion discs is, 
however, still unclear. The quasi-periodic modulation to the X-ray flux could be generated, for instance, 
by matter blobs orbiting in the inner accretion disk at the Lense-Thirring precession frequency~\citep{stella98, cui98}. 
However, also intrinsic disk oscillations of a thin accretion disk~\citep[see][]{wagoner99,kato01}
or a warped disk~\citep{fragile01} may impart a modulation to the spectrum. 
Pointing out the 3:2 ratio between some QPOs in different black hole binaries, 
\citet{abramowicz01} have proposed a resonance model in which these pairs of QPOs are 
excited by the beat between the Keplerian and epicyclic frequencies of a particle orbiting around 
the central compact object. \citet{schnittman04} developed the first model that takes into account general relativistic ray-tracing. 
They considered a hot spot radiating isotropically on nearly circular equatorial orbits around a Kerr black hole. 
\citet{tagger06}  advocate that QPOs in microquasars could be triggered by the Rossby wave instability
in accretion disks surrounding a central compact object. 
For this model ray-traced light curves have recently been developed by~\citet{vincent12}.  

The first study of a QPO model involving thick accretion structure (tori)
was developed by~\citet{rezzolla03} who showed that p-mode oscillation
of a numerically computed accretion torus can generate QPOs. Ray-traced light curves and
power spectra of this model where derived by~\citet{schnittman06}. 
Analytic torus models have the advantage of providing a framework in which all physical quantities 
are accessible throughout the accretion structure and at all times. 
The analysis of these in the context of QPOs has been initiated by~\citet{bursa04} who performed simulations of
ray-traced light curves and power spectra of an optically thin oscillating slender torus. 
This model took only into account simple vertical and radial sinusoidal motion
of a circular cross-section torus (slender tori have nearly circular cross sections). 
In order to allow a more general treatment, a series of theoretical works where dedicated to
developing a proper model of general oscillation modes of a slender
or non-slender perfect-fluid hydrodynamical accretion 
torus~\citep{abramowicz06,blaes06,straub09}. 

This article is a follow-up of the recent analysis of~\citet{mazur13}
who investigated the observable signature of simple time-periodic
deformations of a slender accretion torus. 
This analysis showed that
different deformations lead to very different spectra.
The aim of this article is to go further along the path leading to a fully realistic
model of oscillating tori power spectra. Here, we consider
realistic oscillation modes of slender tori (no longer simple deformations)
as computed by~\citet{blaes06,straub09}. Moreover, all compuations
are performed in the Kerr metric whereas our previous study was restricted
to Schwarzschild black holes.
Our goal is to analyze the spectral signatures of the 
different oscillation modes
and to investigate the effect of the black hole inclination and spin parameters
on the power spectra. 

Section~\ref{sec:equil} describes the equilibrium slender torus while Sect.~\ref{sec:oscil}
describes its oscillations. Sect.~\ref{sec:simu} shows the ray-traced light curves and 
power spectra of the oscillating torus, and Sect.~\ref{sec:conclu} gives conclusions and 
prospects.

%%%%%%%%%	      EQUILIBRIUM 		%%%%%%%%%%%%%%%%%

\section{Equilibrium of a slender accretion torus}
\label{sec:equil}

Spacetime is described by the Kerr metric in the Boyer-Lindquist spherical-like coordinates 
$(t, r, \theta, \varphi)$, with geometrical units $c = 1 = G$, and signature $(-, +, +, +)$. 
The line element has then the following form:
%-----------------------------------------------------------------
\bea
  \mathrm{d}s^2 &=&
  g_{tt} \mathrm{d}t^{2} + 2g_{t\varphi}\mathrm{d}t\mathrm{d}\varphi 
  	+ g_{rr}\mathrm{d}r^2 + g_{\theta\theta}\mathrm{d}\theta^{2} 
		+ g_{\varphi\varphi}\rm{d}\varphi^2Ñ\nn \\
  &=&
  -\left(1-\frac{2 M r}{\Xi}\right) \,\mathrm{d}t^2 
  - \frac{4 M r a }{\Xi} \sin^2\theta\,\mathrm{d}t\, \mathrm{d}\varphi 
  + \frac{\Xi}{\Delta}\, \mathrm{d}r^2  \nn \\
&& 
  + \Xi \,\mathrm{d}\theta^2
  + \left(r^2 + a^2 +\frac{2 M r a^2 \sin^2\theta}{\Xi}\right)\sin^2\theta\, \mathrm{d}\varphi^2,
\label{eq:kerr2}
\index{Kerr metric}
\eea
%-----------------------------------------------------------------
where $M$ is the black hole mass, $a\equiv J/M$ its reduced angular momentum ($J$ being the black hole angular momentum), $\Xi \equiv r^2 + a^2 \cos^2\theta$ and $\Delta \equiv r^2 - 2 M r + a^2$. 

In this spacetime, we consider an axisymmetric, non self-gravitating, perfect fluid, constant
specific angular momentum, circularly orbiting accretion torus. This torus is assumed to be
slender, meaning that its cross section diameter is small compared to its central radius.

%************************************************************
%************************************************************
\subsection{Fluid motion}
%************************************************************
%************************************************************

As the fluid follows circular geodesics, its 4-velocity can be written:
\be
u^{\mu} = A\left( \eta^{\mu}+\Omega \xi^{\mu}\right),
\label{eq-4vel}
\ee
where $\Omega$ is the fluid's angular velocity, and $\eta^{\mu}$ and $\xi^{\mu}$ are the Killing
vectors associated with stationarity and axisymmetry
respectively. The constant $A$ is fixed by imposing the normalization
of the 4-velocity, $u^{\mu}u_{\mu}=-1$.

As spacetime is stationary and axisymmetric, the specific energy $\mathcal{E}$
and specific angular momentum $\mathcal{L}$ defined as:
\bea
\mathcal{E} &=& -\eta^{\mu}u_{\mu} = -u_t,  \\ \nn
\mathcal{L} &=& \xi^{\mu}u_{\mu} = u_\varphi, 
\eea
are geodesic constants of motion.

The rescaled specific angular momentum $\ell$ is defined by
\be
\ell \equiv \frac{\mathcal{L}}{\mathcal{E}} = -\frac{u_{\varphi}}{u_t}.
\ee

We assume this rescaled specific angular momentum
to be constant throughout the torus:
\be
\ell = \ell_0 = {\rm const}.
\ee

The 4-acceleration along a given circular geodesic followed by the fluid is:
\be
a_{\mu} = u^{\nu}\nabla_{\nu}u_{\mu} = -\frac{1}{2\mathcal{U}} \partial_{\mu}\mathcal{U},
\ee
where $\mathcal{U}$ is the effective potential~\citep[see e.g.][]{abramowicz06}:

\be
\mathcal{U} = g^{tt} - 2\ell_0 g^{t\varphi} + \ell_0^{2}g^{\varphi\varphi}.
\ee

The radial and vertical epicyclic frequencies for circular motion are related to 
the second derivatives of this potential:
%------------------------------------------------------------
\begin{eqnarray}
\omega^2_r &=& \frac{1}{2}\frac{{\cal
E}^2}{A^2\,g_{rr}}\left(\frac{\partial^2{\cal U}}{\partial
r^2}\right), \nn \\
\omega^2_{\theta} &=& \frac{1}{2}\frac{{\cal
E}^2}{A^2\,g_{\theta\theta}}\left(\frac{\partial^2{\cal
U}}{\partial \theta^2}\right).
\label{epicyclic-requencies-second-derivatives}
\end{eqnarray}
%------------------------------------------------------------

In the Kerr metric, these epicyclic frequencies at the torus center are: 
%------------------------------------------------------------
\begin{eqnarray}
\omega^2_{r0} &=& \Omega_0^2 \left(1-\frac{6}{r_0}+\frac{8a}{r_0^{3/2}} -\frac{3a^2}{r_0^2}\right), \nn \\
\omega^2_{\theta 0} &=& \Omega_0^2 \left(1-\frac{4a}{r_0^{3/2}} + \frac{3a^2}{r_0^2}\right),
\label{epifreq}
\end{eqnarray}
%------------------------------------------------------------
where a subscript $0$ denotes, here and in the remaining of this article, 
the torus center.

%************************************************************
%************************************************************
\subsection{Fluid thermodynamics}
%************************************************************
%************************************************************

The fluid equation of state is: 
%------------------------------------------------------------
\be
p = K\rho^{(n + 1)/n}, \quad e = np + \rho
\ee
%------------------------------------------------------------
where $p$ is the pressure,
$\rho$ the mass density, $e$ the energy density, $K $ the polytropic
constant and $n$ the polytropic index.

The equation of conservation of energy-momentum leads to the relativistic Euler equation
%------------------------------------------------------------
\begin{equation}
- \frac {\partial_{\mu} p}{p + e} = \partial_{\mu} (\ln {\cal E}),
\label{Euler}
\end{equation}
%------------------------------------------------------------
which has the first integral (Bernoulli equation),
%------------------------------------------------------------
\begin{equation}
H + \ln {\cal E} = {\rm const},
\label{Bernoulli}
\end{equation}
%------------------------------------------------------------
where we have introduced the relativistic enthalpy

\be
H = \int \frac{dp}{p + e}.
\ee

Following~\citet{abramowicz06}, let us define a dimensionless function $f(r, \theta)$ by
%------------------------------------------------------------
\begin{equation}
\frac{p}{\rho} = f(r, \theta)\, \frac{p_0}{\rho_0},
\label{definition-function-f}
\end{equation}
%------------------------------------------------------------
where $p_0$ and $\rho_0$ denote pressure and density at the center
of the torus. \citet{straub09} show that from the
Bernoulli equation (Eq.~\ref{Bernoulli}) one obtains
%------------------------------------------------------------
\begin{equation}
f(r, \theta)= 1 - \frac{1}{nc^2_{s0}}\, \left( \ln {\cal E} - \ln
{\cal E}_0 \right),
\label{general-function-f}
\end{equation}
%------------------------------------------------------------
where $c_{s0}$ is the central value of the sound speed:
\be
c_s^2 = \frac{\partial p}{\partial \rho} = \frac{n + 1}{n}\left(\frac{p}{\rho}\right),
\ee
which is always negligible compared to $1$.

It is easy to derive that
%------------------------------------------------------------
\begin{equation}
\left(\frac{p}{p_0}\right) = f^{n + 1},
~~\left(\frac{\rho}{\rho_0}\right) = f^{n}.
\label{pressure-density-in-terms-of-f}
\end{equation}
%------------------------------------------------------------

The $f$ function takes constant values at isobaric and isodensity surfaces.
In particular it vanishes at the surface of the torus. It will thus be called from
now on {\it surface function}\footnote{{This function is nothing but the Lane-Emden
function for the polytropic fluid.}}.
%************************************************************
%************************************************************
%
\subsection{Surface function}
%
%************************************************************
%************************************************************

Following~\citet{abramowicz06} we introduce a new set of coordinates
%------------------------------------------------------------
\begin{equation}
x = \left( \sqrt{g_{rr}} \right)_0 \left( \frac{r -
r_0}{r_0}\right),~~
y = \left( \sqrt{g_{\theta\theta}} \right)_0 \left( \frac{\pi/2 -
\theta}{r_0}\right).
\label{small-coordinates}
\end{equation}
%------------------------------------------------------------

The dimensionless surface function defined in Eq.~(\ref{general-function-f})
can be expressed using these coordinates
%------------------------------------------------------------
\begin{equation}
f = 1 - \frac{1}{\beta^2}\left(  {\bar \omega}^2_{r0}\, x^2 +
{\bar \omega}^2_{\theta0}\, y^2 \right),
\label{function-f-simple-form}
\end{equation}
%------------------------------------------------------------
where
%------------------------------------------------------------
\begin{equation}
{\bar \omega}_{r0} = \frac{\omega_{r0}}{\Omega_0},\quad
{\bar \omega}_{\theta 0} = \frac{\omega_{\theta 0}}{\Omega_0}
\label{dimensinless-epicycli}
\end{equation}
%------------------------------------------------------------
and
%------------------------------------------------------------
\begin{equation}
\beta^2 = \frac{2n c^2_{s0}}{r^2_0 A^2_0 \Omega^2_0}.
\label{eq:beta}
\end{equation}
%------------------------------------------------------------

As shown by~\citet{abramowicz06}, this parameter is related to the torus
thickness, and the $x$ and $y$ coordinates are of order $\beta$. A torus is slender when

\be
\beta \ll 1.
\ee

We thus define a new set of order-unity coordinates

%------------------------------------------------------------
\begin{equation}
{\bar x} = \frac{x}{\beta}, \quad {\bar y} = \frac{y}{\beta},
\label{magnified-coordinates}
\end{equation}
%------------------------------------------------------------
and
%------------------------------------------------------------
\begin{equation}
f (\xb,\yb)= 1 - \left(  {\bar
\omega}^2_{r0}\, {\bar x}^2 + {\bar \omega}^2_{\theta0}\, {\bar
y}^2 \right).
\label{function-f-simple-form-2}
\end{equation}
%------------------------------------------------------------

The equilibrium slender torus has therefore an elliptical
cross-section (remember that the torus surface is given 
by the zeros of the surface function $f$). 
%Its size is proportional to $\beta$ in small
%coordinates, and it is $\sim 1$ in the ``magnified'' coordinates.

%Note that,
%%------------------------------------------------------------
%\begin{equation}
%%
%\frac{p_0}{\rho_0} = \frac{\beta^2r_0^2 A^2_0 \Omega^2_0}{n + 1}
%%
%\label{small-quantities}
%%
%\end{equation}
%%------------------------------------------------------------

%-----------------------------------------------------------------------

%%%%%%%%%%		 OSCILLATING TORUS		 %%%%%%%%

\section{Oscillations of a slender accretion torus}
%-----------------------------------------------------------------------
%
\label{sec:oscil}

%This Section is devoted to determining the surface function 
%of the torus subject to oscillations at the radial or vertical epicyclic
%frequencies of the equilibrium torus. The 4-velocity of the
%perturbed fluid is also derived.

Following~\citet{straub09} we consider the five lowest-order
oscillation modes with respect to the $\beta$ parameter for slender 
tori (in their article, these modes are labeled from $i=1$ to $i=5$):

\begin{itemize}
\item[1.] radial mode,
\item[2.] vertical mode,
\item[3.] X mode,
\item[4.] plus mode,
\item[5.] breathing mode.
\end{itemize}

This Section is devoted to computing the surface functions
and matter 4-velocity of the slender accretion torus submitted
to these five oscillation modes.

\subsection{Surface functions}

At a given point, the pressure in the oscillating torus
$p^{\prime}$ differs from the pressure in the non-oscillating
torus $p$ by the amount
%------------------------------------------------------------
\begin{equation}
\delta p \propto e^{i(m\varphi - \omega t)},
\label{perturbation-form}
\end{equation}
%------------------------------------------------------------
where $m = 0, 1, 2,...$ is the azimuthal wave number, and $\omega$ is the frequency
of oscillations. The physical pressure perturbation is the real part of the complex quantity above.

It is very useful to express this function in terms
of the following quantity introduced by~\citet{papaloizou84}
%------------------------------------------------------------
\be
W = - \frac{\delta p}{A_0 \rho \Omega_0 \sigma},
\label{eq:Wfunction}
\ee
where the kinematic quantities $A$ and $\Omega$
have been set to their central values, which is a sound
assumption in the slender torus limit. Here $\sigma = \omega / \Omega_0 - m$ is the rescaled mode eigenfrequency.

For any kind of perturbation, the pressure in the oscillating torus, $p^{\prime} = p + \delta p$, 
can be straightforwardly computed:
%------------------------------------------------------------
\be
p^{\prime} = \rho_0\, f^n \left[ \frac{p_0}{\rho_0} f - A_0 \Omega_0 W_i \sigma_i \right],
\ee
where $i$ labels the oscillation mode and $f$ is the unperturbed surface function satisfying $f^{n+1}=p / p_0$.

The expression of the perturbed surface function $f_i^{n+1} = p'/p_0$ corresponding
to oscillation mode $i$ is then to first order in the perturbation
\be
f_i = f - \frac{1}{n+1}\frac{\rho_0}{p_0} A_0 \Omega_0 W_i \sigma_i. 
\ee

\citet{straub09} give the expressions of the $W_i$ functions:
\be
W_i = a_i\,g_i(\xb,\yb)\,\mathrm{cos}\left(m\varphi -(\sigma_i + m)\Omega_0 t\right)
\ee
where $a_i$ are order-unity normalization coefficients and $g_i$ is
a known function of $\xb$ and $\yb$. The choice of normalization
of~\citet{straub09} makes the $W_i$ functions of order unity, which
is not adapted to our present treatment where these functions should
be small corrections. We will thus write the general perturbed surface function
according to

\be
f_i(t,\xb,\yb) = f - \lambda \,\sigma_i \,a_i \,g_i(\xb,\yb)\,\mathrm{cos}\left(m\varphi -(\sigma_i + m)\Omega_0 t\right) 
\label{eq:surfunc}
\ee
where $\lambda$ is a chosen small parameter, and the other quantities appearing in
the second term of the right-hand side are of order unity. The values of $\sigma_i$ are
given in Tab.~2 of~\citet{straub09}, while $a_i\,g_i(\xb,\yb)$ are given in their Tab.~1.

Table~\ref{tab:surffunc} of this article gives the expressions of the surface functions corresponding to
all oscillation modes considered here.

\begin{table*}
\begin{center}
\begin{tabular}{lll}
\hline
Perturbation      & Eigenfrequency $\sigma$      & Surface function   \\
\hline
Radial mode & $\sigma_{r} = {\bar \omega}_{r0}$ & $f  - \lambda \, \sigma_r \,a_r\, {\bar x}\, \mathrm{cos}[m\varphi -(\sigma_{r} + m)\Omega_0 t]$  \\
Vertical mode & $\sigma_{\theta} = {\bar \omega}_{\theta0}$& $f  - \lambda \, \sigma_{\theta}\,a_\theta\,{\bar y}\, \mathrm{cos}[m\varphi -(\sigma_{\theta} + m) \Omega_0 t]$  \\
X mode    & $\sigma_{X} = \sqrt{{\bar \omega}^2_{r0}+{\bar \omega}^2_{\theta0}}$	& $f - \lambda \,\sigma_{X} \, a_X\,{\bar x} {\bar y} \, \mathrm{cos}[m\varphi -(\sigma_X + m) \Omega_0 t]$   \\
Plus mode  &  $\sigma_{+}$  & $f - \lambda \,\sigma_{+}\,a_{+}\,(1+\omega_{41}{\bar x}^2+\omega_{42}{\bar y}^2) \, \mathrm{cos}[m\varphi -(\sigma_+ + m) \Omega_0 t]$    \\
Breathing mode   &	$\sigma_{br}$	  &  $f - \lambda \, \sigma_{br} \,a_{br}\,(1+\omega_{51}{\bar x}^2+\omega_{52}{\bar y}^2) \, \mathrm{cos}[m\varphi -(\sigma_{br} + m) \Omega_0 t]$  \\
\hline
\end{tabular}
\caption{\label{tab:surffunc} {Perturbed surface functions (as defined by Eq.~\ref{eq:surfunc}) for the various kinds of perturbations considered in this work. Here $\lambda$
is a small parameter scaling the perturbation strength. The eigenfrequencies $\sigma_{+}$ and $\sigma_{br}$, as well as the prefactors
$\omega_{ij}$, have complicated expressions, functions of ${\bar \omega}_{r0}$ and ${\bar \omega}_{\theta0}$, that can be found in~\citet{straub09}.
The normalization parameters $a_i$ are of order unity.}}
\end{center}
\end{table*}

%The order of magnitude of surface functions (not taking into account the sinusoidal dependency)
%is a function of three parameters: spin, torus central radius $r_0$ and perturbation parameter $\lambda$
%(see its definition in Table~\ref{tab:surffunc}).
%Let us fix the two last parameters to $r_0=10.8$ and $\lambda=0.1$ (see Section~\ref{sec:param})
%and investigate the impact of spin on surface functions. 
%Fig.~\ref{fig:surfunc} shows the evolution of the surface functions for the radial, minus and plus modes
%in the equatorial plane (i.e. with $\bar{y}=0$) as a function of $\bar{x}$ and for two extreme values of spin
%parameter $a=0$ and $a=0.99$.
%It appears clearly that the minus mode is special in the sense that the
%zero of the surface function (i.e. the location of the torus surface) is depending a lot on the spin value.
%There is even no zero for spin $a=0.99$: this is due to the fact that the torus is responding so much to this mode
%that it is no longer slender, and the above analysis breaks down. On the other hand, all other four modes
%have a much less important dependence on spin value. This is an interesting fact in the perspective of
%studying the impact of the spin parameter on the perturbed torus power spectra.

%\begin{figure*}
%\centering
%\includegraphics[width=0.5\hsize]{FigTorus/Surf0.pdf}
%\includegraphics[width=0.4\hsize]{FigTorus/Surf099.pdf}
%\caption{Equatorial plane surface function for radial and breathing modes.
%	\label{fig:surfunc}}
%\end{figure*}

\subsection{Perturbed fluid 4-velocity}

As shown by~\citet{blaes06}, the fluid 4-velocity is given by the equations
%------------------------------------------------------------
\begin{eqnarray}
\label{four-velocity-general}
&&u^{\mu} = u^{\mu}_0 + \delta u^{\mu},\\ \nn
&&u^{\mu}_0 = A_0(\eta^{\mu} + \Omega_0 \xi^{\mu})\\ \nn
&&\delta u_{\mu} = {\rm Re}\left\{\frac{{\rm i}\rho_0}{p_0 +
e_0}\left(\frac{\partial W}{\partial x^{\mu}}\right)\right\},~~~
\mu = r, \theta
\end{eqnarray}
%------------------------------------------------------------
where $u_0^{\mu}$ is the equilibrium 4-velocity already defined
in Eq.~(\ref{eq-4vel}), which is assumed to be everywhere
equal to its central value, in the slender torus limit.

The expressions of $u_r$ and $u_\theta$ are straightforwardly
derived by using the expression of the $W$ function from~\citet{straub09}.
Table~\ref{tab:4vel} gives these expressions for all oscillation modes.

As the torus is assumed to have a constant rescaled angular momentum $\ell_0$,
the following relation holds between the perturbed temporal and azimuthal
components of the 4-velocity:

\be
u_\varphi = -\ell_0 u_t.
\label{eq:up-ut}
\ee

Adding the normalization condition $u^{\mu}u_\mu=-1$, Eqs.~(\ref{four-velocity-general})
and~(\ref{eq:up-ut}) fully determine the fluid 4-velocity.

\begin{table*}
\begin{center}
\begin{tabular}{lll}
\hline
Perturbation      & $u_r$      & $u_\theta$   \\
\hline
Radial mode & $- \mu\,\sqrt{g_{rr0}} \,a_r\,\mathrm{sin}[m\varphi -(\sigma_{r} + m)\Omega_0 t]$ & $0$  \\
Vertical mode & $0$& $\mu\,\sqrt{g_{\theta\theta0}}\, a_\theta \,\mathrm{sin}[m\varphi -(\sigma_{\theta} + m) \Omega_0 t]$  \\
X mode   & $- \mu \,{\bar y}\,a_X \,\sqrt{g_{rr0}}\,\mathrm{sin}[m\varphi -(\sigma_X + m) \Omega_0 t]$	& $ \mu\, {\bar x} \,a_X\,\sqrt{g_{\theta\theta0}}\,\mathrm{sin}[m\varphi -(\sigma_X + m) \Omega_0 t]$   \\
Plus mode  &  $-2\,\mu\,a_{+}\,\omega_{41}\,{\bar x}\,\sqrt{g_{rr0}}\,\mathrm{sin}[m\varphi -(\sigma_+ + m) \Omega_0 t]$  & $2\,\mu\,a_{+}\,\omega_{42}\,{\bar y}\,\sqrt{g_{\theta\theta0}}\,\mathrm{sin}[m\varphi -(\sigma_+ + m) \Omega_0 t]$     \\
Breathing mode  &  $-2\,\mu\,a_{br}\,\omega_{51}\,{\bar x}\,\sqrt{g_{rr0}}\,\mathrm{sin}[m\varphi -(\sigma_{br} + m) \Omega_0 t]$  & $2\,\mu\,a_{br}\,\omega_{52}\,{\bar y}\,\sqrt{g_{\theta\theta0}}\,\mathrm{sin}[m\varphi -(\sigma_{br} + m) \Omega_0 t]$  \\
\hline
\end{tabular}
\caption{\label{tab:4vel} {Perturbed 4-velocity of the torus matter. Here $\mu$ is a small parameter that scales the perturbation strength.
	It is related to the small parameter $\lambda$ defined in Tab.~\ref{tab:surffunc} by $\mu \approx \lambda \, \sqrt{p_0 / \rho_0}$.}}
\end{center}
\end{table*}

\subsection{Perturbations parameters}
\label{sec:param}

%The parameters of our model are summarized in Tab.~\ref{tab:param}.

The torus central radius is fixed at the value that ensures the condition
$\omega_{\theta 0} / \omega_{r 0}=3/2$ in Schwarzschild metric. This translates
to $r_0 = 10.8$. The torus central
density and polytropic index are fixed at $\rho_0=1$, $n=1.5$. The polytropic
constant is fixed at a value $K=10^{-4}$ that allows to get $\beta \ll 1$ (see
Eq.~\ref{eq:beta}). 

The perturbation strengths $\lambda$ and $\mu$ defined in Tab.~\ref{tab:surffunc}
and~\ref{tab:4vel} can be shown to be related by the simple expression $\mu \approx \lambda \sqrt{p_0 / \rho_0}$.
There is thus only one parameter to choose, and we fix $\lambda = 0.2$ throughout this article.
This value is the largest one that allows the torus to remain slender for
all times and all modes.

The mode number $m$ is fixed to $0$. This results in the torus being
always axially symmetric and allows to compute easily its cross-section
circumference as a function of time, as will be needed to compute its emission
(see Sect~\ref{sec:raytrace}).

The only free parameters that we will vary in this study for all
different oscillation modes are:
\begin{itemize}
\item the black hole spin,
\item the inclination angle.
\end{itemize}

The ranges of variation of all parameters are given in Tab.~\ref{tab:param}.

\begin{table*}
\begin{center}
\begin{tabular}{lll}
\hline
Parameter            & Notation & Value  \\
\hline
Black hole spin  & $a$  & $\{0;0.99\}$\\
Inclination angle  & $i$  & $\{5^{\circ};45^{\circ};85^{\circ}\}$ \\
Torus central radius    	& $r_0$  & $10.8$  \\
Polytropic constant     & $K$   & $10^{-4}$ \\
Polytropic index     		  &  $n$ & $1.5$ \\
Central density    		  &  $\rho_0$ & $1$ \\
Mode number   & $m$ &   $0$ \\
Perturbation strength   & $\lambda$ &  $0.2$\\
\hline
\end{tabular}
\caption{\label{tab:param} {Values of parameters that describe our model for all five modes of oscillation.}}
\end{center}
\end{table*}

%

%%%%%%%%%%%		RAY-TRACING SIMU	     	%%%%%%%%%%%%%%%%%

%-----------------------------------------------------------------------
\section{Light curves and power spectra of the oscillating torus}
%-----------------------------------------------------------------------
\label{sec:simu}

\subsection{Ray-tracing the oscillating torus}
\label{sec:raytrace}

Light curves and power spectra of the oscillating tori are computed
by using the open-source
general relativistic ray-tracing code \texttt{GYOTO}\footnote{Freely available at \url{gyoto.obspm.fr}}\citep{vincent11}.

Null geodesics are integrated backward in time from a distant observer's screen 
to the optically thick torus. The zero of the surface function is found numerically 
along the integrated geodesic and the fluid 4-velocity is determined by using 
Eqs.~(\ref{four-velocity-general}).

The torus is assumed to emit radiation isotropically in its rest frame.
Among the five different oscillation modes investigated here,
only three lead to a variation of the emitting surface:
the radial mode and the plus and breathing modes.
The radial mode leads to a change in $r_0$, with constant 
cross-section circumference $\mathcal{C}$. 
The plus and breathing modes give rise to a change in $\mathcal{C}$, 
with constant $r_0$. The emitting surface area
being proportional to $r_0 \times \mathcal{C}$ (as the azimuthal
wavenumber $m=0$) these three
modes lead to a change of emitting surface area.
In order to take into account in the simplest way the
resulting change in observed flux, we assume
that the emitted intensity at the torus surface is

\be
I_{\nu} = \left\{
\begin{array}{l}
 1, \,\qquad\qquad\qquad\qquad\:\:\,\mathrm{vertical, X\: modes}, \vspace{0.2cm}\\
 r_0^{\mathrm{eq}} \mathcal{C}^{\mathrm{eq}}/ r_0(t_{\mathrm{em}}) \mathcal{C}(t_{\mathrm{em}}), \,\mathrm{radial, plus, breathing\: modes} \\
\end{array}
\right.
\label{eq:Iem}
\ee
where $t_{\mathrm{em}}$ is the time of emission and the superscript 
refers to equilibrium torus quantities.
This choice boils down to assuming that the torus
luminosity is constant.
We have computed numerically the quantities $r_0(t)$
and $\mathcal{C}(t)$
as a function of time for all three modes, 
and interpolate to determine their values
at any emission time.

Maps of specific intensity are computed with a resolution of 
$1000\times1000$ pixels. The time resolution is of around $20$ snapshots
per Keplerian period at the central radius of the torus. This choice
is a balance between the necessity to use a high enough time 
resolution in order to compute accurate power spectral densities (PSD)
and the computing time needed for the simulation. One light curve
lasts in total $4.5$ Keplerian orbits at $r_0$, divided in
$100$ snapshots, and needs a total computing time corresponding to $75$~h
for one CPU.

\subsection{Power spectra of torus oscillations}

Power spectral densities are computed in the following way.
Let $\delta t$ be the time step between two consecutively computed maps
of specific intensity and $N$ be the total number of calculated maps.
Let $F(t_k)$ be the observed flux at observation time $t_k = k\delta t$.
$F(t_k)$ is simply the sum over all pixels of the intensity map.
The power spectral density at frequency $f_k=k/ (N \delta t)$
is defined as the square of the modulus
of the fast Fourier transform of the signal, thus:

\be
PSD(f_k)= \left| \sum_{j=0}^{N-1} F(t_j) \mathrm{exp}(2\pi i j k/N)\right|^{2}
\ee

\subsection{Results and discussion}

\begin{figure*}
\begin{minipage}{0.9\hsize}
\includegraphics[width=1\hsize]{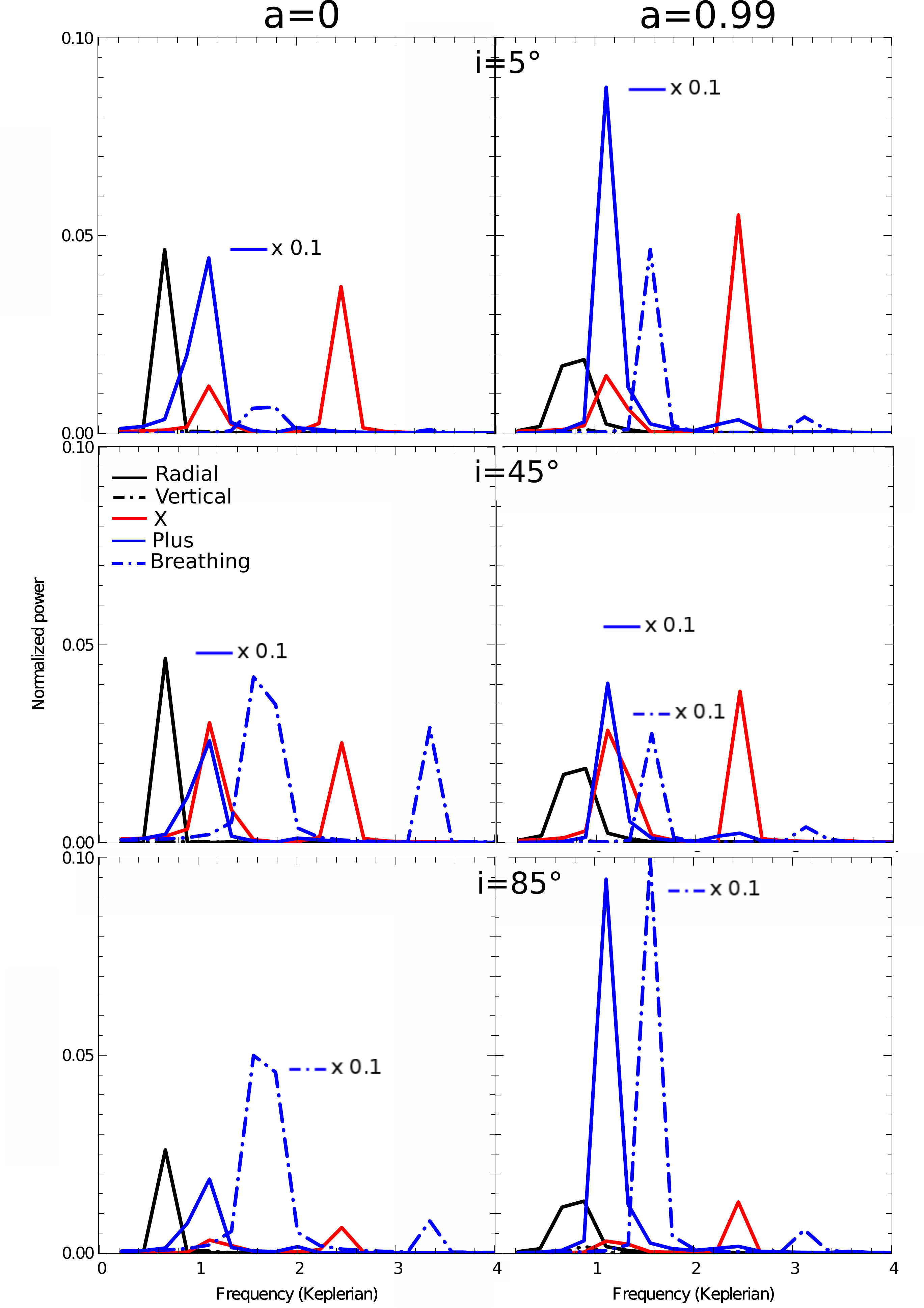}
\end{minipage}
\caption{Power spectral density of a slender oscillating torus for spin $0$ (left column) or $0.99$ (right column)
	and inclination $5^{\circ}$ (upper row), $45^{\circ}$ (middle row) or $85^{\circ}$ (lower row).
	All five oscillation modes investigated in this article are represented on each panel (radial mode in solid black,
	vertical mode in dashed black -- {invisible because too weak}, X mode in solid red, plus mode in solid blue and breathing mode in dashed blue).
	Note that some spectra are multiplied by $0.1$ for convenience.
	Frequencies are in units of the Keplerian frequency at $r_0$ for spin $0$.} \label{fig:allPSD}
\end{figure*}

This Section shows the light curves and PSD of all five torus oscillation modes 
for two extreme values of spin, $a=0$ and $a=0.99$, and
three values of inclinations $i=5^{\circ}$, $i=45^{\circ}$ and $i=85^{\circ}$.
All other parameters are fixed to the values given in Tab.~\ref{tab:param}.
Figs.~\ref{fig:allLC} and~\ref{fig:allPSD} show respectively the light curves 
and PSD for all oscillation modes, spins and inclinations.
Three effects are particularly important to explain the modulation
of light curves and the corresponding amplitude of PSD:
\bi
\item the general relativistic redshift effect that has an impact on the observed
intensity: $I_\nu^{\mathrm{obs}} = g^4 I_\nu^{\mathrm{em}}$
where $g=\nu^{\mathrm{em}} / \nu^{\mathrm{obs}}$ and $\nu^{\mathrm{em}}$ is related to
the emitter's 4-velocity,
\item the time variation of the emitting area in the emitter's frame $A_{\mathrm{em}}$, that
is linked to the variation of the surface function, and the resulting change of
emitted intensity (see Eq.~\ref{eq:Iem}),
\item the time variation of the apparent torus area on the
observer's sky $A_{\mathrm{app}}$, that is a function of the variation
of $A_{\mathrm{em}}$, of the bending of photons geodesics and of 
the inclination angle.
\ei
%In particular, even if the emitting area $A_{\mathrm{em}}$ is constant
%in time, the apparent area $A_{\mathrm{app}}$ may vary, depending
%on the inclination.

The vertical mode is relevant to determine the impact of the first effect.
This mode has constant $A_{\mathrm{em}}$, and $A_{\mathrm{app}}$
is also barely varying for all inclinations. The light curve modulation is
thus mainly due to the variation of the emitter's 4-velocity. Given the
extremely low level of modulation for this vertical mode, it appears
that the effect of general relativistic redshift variation is negligible
compared to the time variation of the emitting and the apparent torus areas.

The observed flux is essentially equal to the apparent area $A_{\mathrm{app}}$ of the torus
multiplied by the observed intensity. Neglecting the redshift effect, the observed intensity varies
due to the intrinsic variation of the emitted intensity which in turn is linked to
the intrinsic variation of the emitted area $A_{\mathrm{em}}$. The total
modulation depends on the respective weights of these effects. 

\subsubsection{Impact of inclination}

The very
strong dependence of the spectra on inclination shows that the most
important effect is the variation of apparent area, which is the only
inclination dependent effect.
This can be quantified by comparing the
variation of $A_{\mathrm{em}}$ and $A_{\mathrm{app}}$. Let us take the breathing
mode at spin $0$ as an example. 
While the emitting area can vary up to $20\%$ during the course of the oscillation, 
the apparent area changes by $12\%$ for a face-on torus ($i=5^{\circ}$) and by $80\%$ 
for and edge-on torus ($i=85^{\circ}$), whereas the flux varies by $6\%$ at $5^{\circ}$ 
and $65\%$ at $85^{\circ}$, respectively.
The variation of $A_{\mathrm{em}}$ (and subsequent inverse variation of
emitted intensity) thus leads to somewhat restraining the effect of the
variation of $A_{\mathrm{app}}$. 
\begin{figure*}
\centering
\includegraphics[width=0.45\hsize]{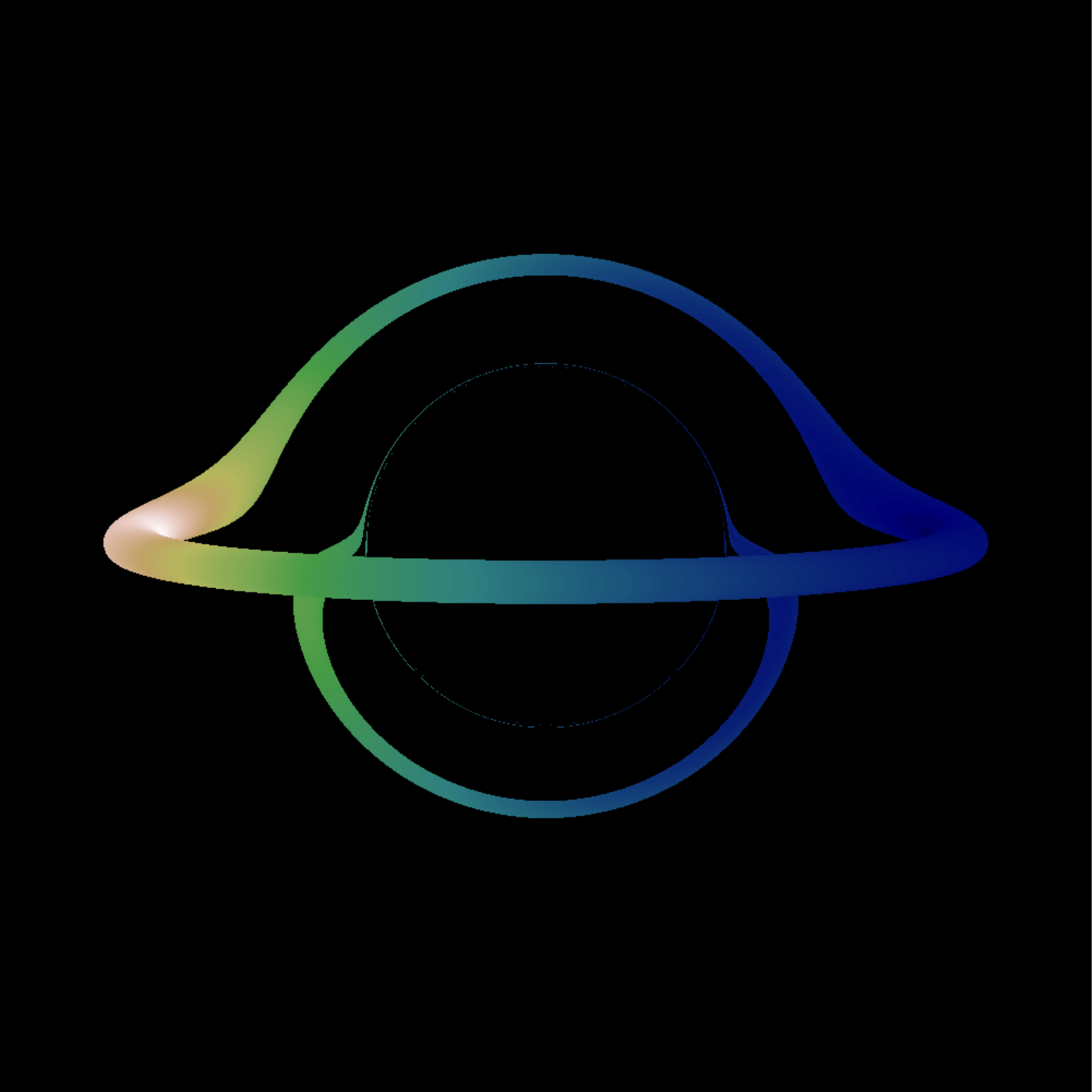}
\includegraphics[width=0.45\hsize]{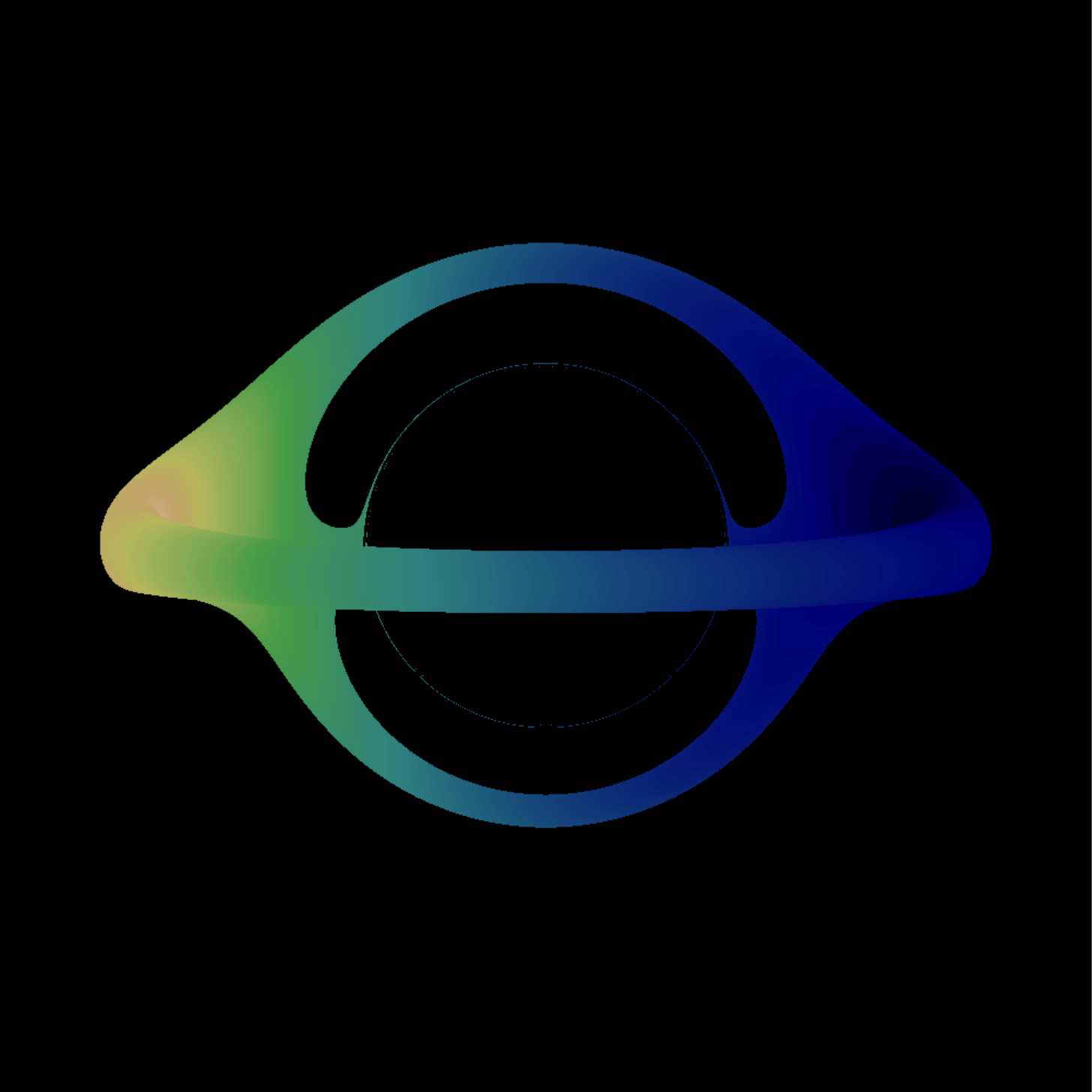}
\caption{Breathing mode images at spin $0$ and inclination $85^{\circ}$. The two panels
	correspond to the two extreme values of the torus apparent area. The observed 
	intensity of every pixel is reduced when the apparent size is bigger, as the emitted
	intensity decreases when the emitting area increases.
	Although it does not seem so from the torus apparent area variation, the torus
	remains slender at all times. These images are highly distorted by general relativistic
	effects. 
	} \label{fig:plusmode}
\end{figure*}
Fig.~\ref{fig:plusmode} shows the two images
of the breathing mode at $85^{\circ}$ of inclination corresponding to the extreme 
values of $A_{\mathrm{app}}$.

The plus and breathing modes are particularly sensitive to inclination. This
is due to the fact that the torus cross-section is modified differently
in the vertical and horizontal directions, where vertical refers to
the direction orthogonal to the equatorial plane and horizontal
inside the equatorial plane. The plus mode leads mainly to a horizontal motion 
whereas the breathing mode induces mainly a vertical movement.
This ends up in the plus mode being more modulated at small inclinations 
as horizontal motions are more visible for a face-on observer.
The opposite
is true for the breathing mode.

The X mode is special in the sense that it is the only mode
leading to a strong first harmonic whatever the inclination.
This is due to the particular variation of the torus surface for
this mode, that leads to two local maxima of apparent area $A_{\mathrm{app}}$
per Keplerian period. These two local maxima of $A_{\mathrm{app}}$ have different values, 
leading to different values of flux local maxima.

%These dependencies of the spectra on the inclination parameter are
%the same at both spins. Let us now investigate the differences between 
%spectra at spin $0$ and spin $0.99$.

\subsubsection{Impact of spin}

\begin{figure*}
\begin{minipage}{0.9\hsize}
\includegraphics[width=1\hsize]{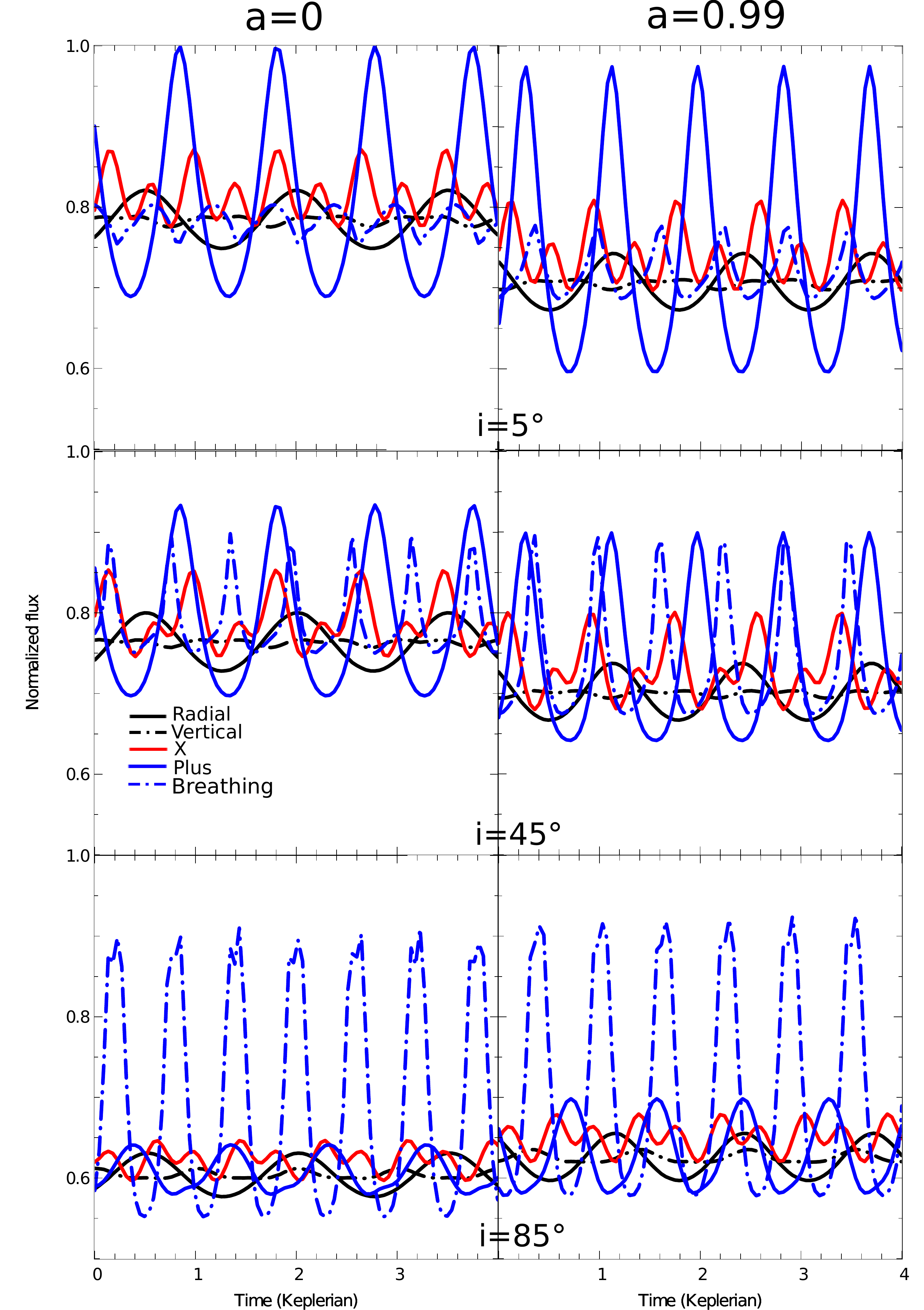}
\end{minipage}
\caption{Light curves of a slender oscillating torus for spin $0$ (left column) or $0.99$ (right column)
	and inclination $5^{\circ}$ (upper row), $45^{\circ}$ (middle row) or $85^{\circ}$ (lower row).
	All five oscillation modes investigated in this article are represented on each panel (radial mode in solid black,
	vertical mode in dashed black, X mode in solid red, plus mode in solid blue and breathing mode in dashed blue).
	Times are in units of the Keplerian period at $r_0$ for spin $0$.} \label{fig:allLC}
\end{figure*}

The first difference due to the spin parameter is the value of the peak
frequency, $\sigma_i$ (see Tab.~\ref{tab:surffunc}), which is a function
of spin. For instance, $\sigma_+$ varies from $1.03$ ($a=0$) to $1.2$ ($a=0.99$)
in units of the Keplerian frequency. The breathing mode
frequency varies less, going from $1.67$ ($a=0$) to $1.64$ ($a=0.99$).
This spin-related difference of frequency is clearly visible in the upper and
lower panels of Fig.~\ref{fig:allLC}.

Regarding the total power of the modulation, the most stringent difference
is associated to the plus and breathing modes that always lead to a high-spin power 
greater or equal to the low-spin power.
For instance, the plus mode at inclination
$5^{\circ}$ has around $40\%$ more power at spin $0.99$ than at spin $0$.
This increase in power with spin is dictated by the combined impact of the
variation of emitting and apparent areas. 
Taking the plus mode as an example, its emitting area 
varies a bit less at higher spin while the apparent area 
at low inclination varies significantly more. 
These two effects lead
to a higher modulation at higher spin.

Modes other than plus and breathing have comparable power at both spins due
to the fact that the apparent area of the torus varies by a similar amount
for the two spin values.

A third effect of the spin parameter is to change the relative values
of the powers of different modes. For instance, the power ratio between
the plus and breathing modes at inclination $5^{\circ}$ is of around $50$ at spin $0$
and $20$ at spin $0.99$. At inclination $85^{\circ}$ this ratio is evolving
from $35$ to $10$. The dependence of this ratio with spin is mainly dictated by
the evolution of the apparent area: the respective evolutions of $A_{\mathrm{app}}$ 
for the plus and breathing modes are more different at lower than at higher spin.

\vspace{0.5cm}

The conclusion of this analysis is that both inclination and spin have
a clear impact on the power spectra. These parameters are key to
determining the apparent area of the torus as a function of time, which
itself dictates the modulation of the light curve. 
In particular, the plus and breathing modes
are extremely sensitive to these two parameters: 
\bi
\item The peak frequency of their PSD is a function of spin;
\item Plus mode has
less power at higher inclination and breathing mode more power;
\item Plus and breathing modes have more power at higher spins;
\item The ratio of the breathing mode to plus mode powers is a function
of spin and inclination.
\ei
These features may be used to constrain spin and inclination
by fitting the theoretical spectra to observed QPO data. However, 
our model is still too simple to allow us putting this possibility
to the test.

%%%%%%%%%%%		CONCLUSION	     	%%%%%%%%%%%%%%%%%

%-----------------------------------------------------------------------
\section{Conclusions and perspectives}
%-----------------------------------------------------------------------
\label{sec:conclu}

This article is devoted to computing the spectral signature
of the lowest order oscillation modes of slender accretion
tori with constant angular momentum surrounding Kerr
black holes.

We show how the ray-traced light curves and power spectral densities are regulated by the oscillation mode, 
the inclination and the spin parameter. The power of a given mode has a strong and non-linear dependence on spin and inclination. 
We find that both plus and breathing modes dominate the high spin regime at all inclinations.
For low spin, either the plus or the breathing mode is dominating, with a strong dependence
on the inclination parameter. Lower order modes are much less sensitive to changing the spin,
but they still vary with the inclination.
This may imply that depending on black hole spin and inclination of an X-ray binary system the observed 
high-frequency QPOs would be attributed to different oscillation modes. If true, then there would be no unique 
pair of modes that describes the observed 3:2 resonance in all black hole sources with QPOs. Rather, 
each black hole binary system would have its own characteristic mode combination which would be visible at its 
individual inclination and spin. However, we caution that our model is still too simple to draw any conclusions. 
How QPOs are excited and which oscillation modes are at work is at this point still unclear. 
Magneto-hydrodynamical simulations of radiation dominated accretion discs suggest on the one hand that the 
breathing mode is the strongest of the oscillation modes, but on the other hand that it is, 
like all higher order acoustic modes (but unlike the epicyclic modes), damped at a 
rate proportional to the disc scale height~\citep[see][]{blaes11}.

{This work also highlights the importance of ray-tracing for computing power spectra of oscillating
tori. Previous works have suggested that the 3:2 double peak may be due to a resonance between
the radial and vertical modes~\citep{abramowicz01}, the radial and plus modes~\citep{rezzolla03}
or the vertical and breathing modes~\citep{blaes06}. The fact that the power exhibited by these modes
is so different in our simulations (which is a consequence of ray-tracing, as the power is mainly
linked with the variation of the torus apparent area) should be taken into account in these models.}

%We show how these spectra depend on
%the oscillation mode, on the inclination parameter, and on the spin
%parameter. The dependency on the inclination and spin
%parameters being very strong, it may lead
%to observable differences.

A natural follow-up of this analysis will be to 
determine to what extent inclination and spin could be constrained
by fitting this model to QPO data. 
In order to be able to answer this question, the oscillating torus
model should be made more realistic. Future work will be devoted
to developing general-relativistic magneto-hydrodynamics (GRMHD)
simulations of non-equilibrium slender tori in order to determine
which oscillation modes are the most likely to be triggered and with
what strength. 
%Knowing this result will allow us to restrict
%our PSD computations to the modes that are actually triggered
%in accretion tori, and fit these modes to QPO data.

\newpage

\begin{acknowledgements}
We acknowledge support from the Polish NCN
UMO-2011/01/B/ST9/05439 grant, 
from the Polish NCN grant 2013/08/A/ST9/00795
and the Swedish VR grant.
Computing was partly done using the Division Informatique de
l'Observatoire (DIO) HPC facilities from Observatoire de Paris
(\url{http://dio.obspm.fr/Calcul/}).
GT and PB acknowledge the Czech research grant GA\v{C}R~209/12/P740 and the project CZ.1.07/2.3.00/20.0071 "Synergy" aimed to support international collaboration of the Institute of physics in Opava.
\end{acknowledgements}

\bibliographystyle{aa}
\bibliography{OscillatingTorus}

\end{document}